\UseRawInputEncoding
\documentclass[pra,twocolumn,shownopacs,amssymb]{revtex4}
\usepackage{graphicx,psfrag,amsmath,amssymb}
    
\begin{document}

\title{Two-body problem in a multiband lattice and the role of quantum geometry}

\author{M. Iskin}
\affiliation{Department of Physics, Ko\c{c} University, Rumelifeneri Yolu, 
34450 Sar\i yer, Istanbul, Turkey}

\date{\today}

\begin{abstract}

We consider the two-body problem in a periodic potential, and study the bound-state 
dispersion of a spin-$\uparrow$ fermion that is interacting with a spin-$\downarrow$ 
fermion through a short-range attractive interaction. Based on a variational approach, 
we obtain the exact solution of the dispersion in the form of a set of self-consistency 
equations, and apply it to tight-binding Hamiltonians with onsite interactions. 
We pay special attention to the bipartite lattices with a two-point basis that exhibit 
time-reversal symmetry, and show that the lowest-energy bound states disperse 
quadratically with momentum, whose effective-mass tensor is partially controlled 
by the quantum metric tensor of the underlying Bloch states. In particular, we apply our 
theory to the Mielke checkerboard lattice, and study the special role played by 
the interband processes in producing a finite effective mass for the bound states 
in a non-isolated flat band.

\end{abstract}

\maketitle

\section{Introduction}
\label{sec:intro}

A flat band refers to a featureless Bloch band in which the energy of a single 
particle does not change when the crystal momentum is varied across the 1st 
Brillouin zone. Because of their peculiar properties~\cite{balents20, liu14, 
leykam18, tasaki98, parameswaran13}, there is a growing demand in designing 
and studying physical systems that exhibit flat bands in their 
spectrum~\cite{jo12, nakata12, li18, diebel16, kajiwara16, ozawa17}. 
For instance, such a dispersionless band indicates that 
not only the effective mass of the particle is literally infinite but also its group 
velocity is zero. This further suggests that the particle remains localized in real 
space. Then, up until very recently~\cite{peotta15}, one of the puzzling questions 
was whether the diverging effective mass is good or bad news for the fate of 
superconductivity in a material that is to a large extent characterized by a flat
band, given that superconductivity, by definition, requires a finite effective mass 
for its superfluid carriers.

Despite such a complicacy that prevents the motion of particles through the intraband 
processes in a flat band, it turns out that the superfluidity of many-body bound states 
is still possible through the interaction-induced interband transitions in the presence 
of other flat and/or dispersive bands~\cite{peotta15}. Furthermore, in the case of 
an isolated flat band, i.e., a flat band that is separated by some energy gaps 
from the other bands, it has been shown that the effective mass of the two-body 
bound states becomes finite as soon as the attractive interaction between the 
particles is turned on, independently of its strength~\cite{torma18}. Moreover, 
assuming that the interaction is weak, the effective-mass tensor is characterized 
by the summation of the so-called quantum-metric tensor~\cite{provost80, berry89, 
resta11} of the flat band in the 1st Brillouin zone. There is no doubt that such 
few-body problems offer a bottom-up approach for the analysis of the many-body 
problem, e.g., it may be possible to use the two-body problem as a universal 
precursor of superconductivity in a flat band~\cite{torma18}.

Motivated also by related proposals in other contexts~\cite{iskin18a, wang20}, 
here we construct a variational approach to study the two-body bound-state 
problem in a generic multi-band lattice, and give a detailed account of bipartite 
lattices with a two-point basis and an onsite interaction that manifest time-reversal 
symmetry. For this case, we show that the lowest-energy bound states 
disperse quadratically with momentum, whose effective-mass tensor has 
two physically distinct contributions coming from (i) the intraband processes 
that depend only on the one-body dispersion and (ii) the interband processes 
that also depend on the quantum-metric tensor of the underlying Bloch states. 
In particular we apply our theory to the Mielke checkerboard lattice for its 
simplicity~\cite{iskin19b}, and reveal how the interband processes help produce 
a finite effective mass for the bound states in a non-isolated flat band, i.e., a flat 
band that is in touch with others. Recent realizations of non-isolated flat bands 
include the Kagome and Lieb lattices~\cite{jo12, nakata12, li18, 
diebel16, kajiwara16, ozawa17}, but they both involve a relatively complicated 
three-point basis.

The remaining parts of this paper are organized as follows. In Sec.~\ref{sec:va} 
we introduce the two-body Hamiltonian for a general multi-band lattice, and 
present its bound-state solutions through a variational approach. In Sec.~\ref{sec:tp}
we focus on the tight-binding lattices with a two-point basis, and derive their 
self-consistency equations in the presence of a time-reversal symmetry. 
In Sec.~\ref{sec:num} we analyze the bound-state problem in a non-isolated 
flat band, and discuss the role of quantum metric. In Sec.~\ref{sec:conc} we 
end the paper with a brief summary of our conclusions.

\section{Variational Approach}
\label{sec:va}

In this paper we are interested in the dispersion of the two-body bound-state in a periodic 
potential when a spin-$\uparrow$ fermion interacts with a spin-$\downarrow$ fermion 
through a short-range attractive interaction~\cite{torma18, ohashi08}. Our starting Hamiltonian 
can be written as 
$
H = H_0 + H_{\uparrow\downarrow},
$
where the one-body contributions $H_0 = \sum_\sigma H_\sigma$ are governed by
\begin{align}
\label{eqn:H0}
H_\sigma = \int d\mathbf{x} \psi_\sigma^\dagger(\mathbf{x})
\left[- \frac{\nabla^2}{2m_\sigma} + V_\sigma (\mathbf{x}) \right]
\psi_\sigma(\mathbf{x}).
\end{align}
Here the operator $\psi_\sigma(\mathbf{x})$ annihilates a spin-$\sigma$ fermion
at position $\mathbf{x}$, the Planck constant $\hbar$ is set to unity, and 
$V_\sigma (\mathbf{x})$ is the periodic one-body potential. Without loss of generality, 
the one-body problem can be expressed as
\begin{align}
\label{eqn:Hsigma}
H_\sigma | n \mathbf{k} \sigma \rangle 
= \varepsilon_{n \mathbf{k} \sigma} | n \mathbf{k} \sigma \rangle,
\end{align}
where $| n \mathbf{k} \sigma \rangle$ represents a particle in the Bloch state that is 
labeled by the band index $n$ and crystal momentum $\mathbf{k}$ in the 1st Brillouin 
zone, and $\varepsilon_{n \mathbf{k} \sigma}$ is the corresponding one-body dispersion. 
The Bloch wave function can be conveniently chosen as
$
\phi_{n \mathbf{k} \sigma} (\mathbf{x}) = \langle \mathbf{x} | n \mathbf{k} \sigma \rangle 
= e^{i \mathbf{k} \cdot \mathbf{x}} n_{\mathbf{k} \sigma}(\mathbf{x}) / \sqrt{N_c},
$ 
where $n_{\mathbf{k} \sigma}(\mathbf{x})$ is a periodic function in space and 
$N_c$ is the number of unit cells in the system. 
We note that if $N_b$ is the number of basis sites in a unit cell, i.e., the number of sublattices
in the system, then the total number of lattice sites is $N = N_b N_c$, and 
$
\int d\mathbf{x} = N_c \int_\textrm{unitcell} d \mathbf{x}.
$

The two-body contribution to the Hamiltonian can be written in general as
\begin{align}
\label{eqn:Hupdown}
H_{\uparrow\downarrow} = \int d\mathbf{x}_1 d\mathbf{x}_2 \psi_\uparrow^\dagger(\mathbf{x}_1)
\psi_\downarrow^\dagger(\mathbf{x}_2) U (\mathbf{x}_{12})
\psi_\downarrow(\mathbf{x}_2) \psi_\uparrow(\mathbf{x}_1),
\end{align}
where the two-body potential $U (\mathbf{x}_{12})$ depends on the relative 
position $\mathbf{x}_{12} = \mathbf{x}_1-\mathbf{x}_2$ of the particles 
and has the same periodicity as the one-body potentials. 
It is convenient to express $H_{\uparrow\downarrow}$ in terms of the Bloch wave functions.
For this purpose, we combine the Fourier expansions of the Bloch state
$
| n \mathbf{k} \sigma \rangle = \frac{1}{\sqrt{N_c}} \sum_j 
e^{i \mathbf{k} \cdot \mathbf{x}_j} | n j \sigma \rangle,
$
where $\mathbf{x}_j$ is the position of the lattice site $j$, and the Wannier function
$
W_{n \sigma} (\mathbf{x} - \mathbf{x}_j) = \frac{1}{\sqrt{N_c}} \sum_\mathbf{k} 
e^{- i \mathbf{k} \cdot \mathbf{x}_j} \phi_{n\mathbf{k} \sigma}(\mathbf{x}),
$
where
$
W_{n \sigma} (\mathbf{x} - \mathbf{x}_j) = \langle \mathbf{x} | n j \sigma \rangle
$
is the usual definition in the tight-binding approximation. This leads to
$
| \mathbf{x} \sigma \rangle = \sum_{n j} W_{n \sigma}^*(\mathbf{x} - \mathbf{x}_j) |n j \sigma \rangle,
$
suggesting that
\begin{align}
\label{eqn:psisigma}
\psi_\sigma(\mathbf{x}) = \sum_{n \mathbf{k}} \phi_{n \mathbf{k} \sigma} (\mathbf{x}) 
c_{n \mathbf{k} \sigma}.
\end{align}
Here the operator $c_{n \mathbf{k} \sigma}$ annihilates a spin-$\sigma$ fermion in the 
$n$th Bloch band with momentum $\mathbf{k}$. 

The two-body dispersion $E_\mathbf{q}$ is determined by the Schr\"odinger equation
\begin{align}
\label{eqn:H}
H | \Psi_\mathbf{q} \rangle = E_\mathbf{q} | \Psi_\mathbf{q} \rangle,
\end{align}
where $\mathbf{q}$ is the total momentum of the particles and $| \Psi_\mathbf{q} \rangle$
represents the two-body bound state for a given $\mathbf{q}$. Here the conservation 
of $\mathbf{q}$ is due to the discrete translational invariance of $H$. The exact solutions 
of $E_\mathbf{q}$ can be achieved by the functional minimization of
$
\langle \Psi_\mathbf{q} | H - E_\mathbf{q} | \Psi_\mathbf{q} \rangle
$
~\cite{torma18, ohashi08}, where 
\begin{align}
\label{eqn:Psiq}
| \Psi_\mathbf{q} \rangle = \sum_{nm\mathbf{k}} \alpha_{nm\mathbf{k}}^\mathbf{q} 
c_{n, \mathbf{k}+\frac{\mathbf{q}}{2}, \uparrow}^\dagger
c_{m, -\mathbf{k}+\frac{\mathbf{q}}{2}, \downarrow}^\dagger | 0 \rangle
\end{align}
is the most general variational ansatz (i.e., for a given $\mathbf{q}$) with complex parameters 
$\alpha_{nm\mathbf{k}}^\mathbf{q}$. 
Here $| 0 \rangle$ represents the vacuum of particles and the normalization of 
$| \Psi_\mathbf{q} \rangle$ requires 
$
\sum_{nm\mathbf{k}} |\alpha_{nm\mathbf{k}}^\mathbf{q}|^2 = 1.
$
Unlike the continuum model of uniform systems where the bound-state wave function
involves pairs of particles with $\mathbf{k}+\frac{\mathbf{q}}{2}$ and $-\mathbf{k}+\frac{\mathbf{q}}{2}$
within a single parabolic band, here we also allow $n \ne m$ terms to take the interband 
couplings that are induced by the periodic lattice potential into account. 
They correspond to pairs of particles whose center-of-mass momenta are shifted by 
reciprocal-lattice vectors in the extended-zone scheme~\cite{ohashi08}.
By plugging Eq.~(\ref{eqn:psisigma}) in Eq.~(\ref{eqn:Hupdown}), a compact way 
to present the functional is
\begin{align}
\label{eqn:functional}
\langle H - E_\mathbf{q} \rangle &= \sum_{nm\mathbf{k}}  
(\varepsilon_{n, \mathbf{k}+\frac{\mathbf{q}}{2}, \uparrow} 
+ \varepsilon_{m, -\mathbf{k}+\frac{\mathbf{q}}{2}, \downarrow} - E_\mathbf{q}) 
|\alpha_{nm\mathbf{k}}^\mathbf{q}|^2 \nonumber \\
&+ \frac{1}{N_c} \sum_{nm n'm'; \mathbf{k} \mathbf{k'}}  
U_{n'm'\mathbf{k'}}^{nm\mathbf{k}}(\mathbf{q}) \alpha_{n'm'\mathbf{k'}}^{\mathbf{q}*} \alpha_{nm\mathbf{k}}^\mathbf{q},
\end{align}
where the non-interacting terms are simply determined by Eq.~(\ref{eqn:Hsigma})
and the most general interaction-dependent matrix elements are given by a 
complicated integral
\begin{align}
\label{eqn:Ugen}
U_{n'm'\mathbf{k'}}^{nm\mathbf{k}} & (\mathbf{q}) 
= \frac{1}{N_c} \int d\mathbf{x}_1 d\mathbf{x}_2 
{n'}_{\mathbf{k'}+\frac{\mathbf{q}}{2},\uparrow}^*(\mathbf{x}_1) 
{m'}_{-\mathbf{k'}+\frac{\mathbf{q}}{2},\downarrow}^*(\mathbf{x}_2)  \nonumber \\
\times & U(\mathbf{x}_{12})
e^{i(\mathbf{k}-\mathbf{k'}) \cdot \mathbf{x}_{12}}
m_{-\mathbf{k}+\frac{\mathbf{q}}{2},\downarrow}(\mathbf{x}_2)
n_{\mathbf{k}+\frac{\mathbf{q}}{2},\uparrow}(\mathbf{x}_1).
\end{align}
Then we set 
$
\partial \langle H - E_\mathbf{q} \rangle / \partial \alpha_{nm\mathbf{k}}^{\mathbf{q}*} = 0,
$ 
and obtain an integral equation that must be self-consistently satisfied by both 
$\alpha_{nm\mathbf{k}}^\mathbf{q}$ and $E_\mathbf{q}$ as
\begin{align}
\label{eqn:alphagen}
\alpha_{nm\mathbf{k}}^\mathbf{q} = - \frac{ \frac{1}{N_c} 
\sum_{n'm'\mathbf{k'}} U_{n m \mathbf{k}}^{n'm'\mathbf{k'}} \alpha_{n'm'\mathbf{k'}}^\mathbf{q}}
{\varepsilon_{n, \mathbf{k}+\frac{\mathbf{q}}{2}, \uparrow} 
+ \varepsilon_{m, -\mathbf{k}+\frac{\mathbf{q}}{2}, \downarrow} - E_\mathbf{q}}.
\end{align}
To simplify Eqs.~(\ref{eqn:Ugen}) and~(\ref{eqn:alphagen}), next we restrict our analysis 
to the zero-ranged contact interactions where
$
U(\mathbf{x}_{12}) = U(\mathbf{x}_1) \delta(\mathbf{x}_{12})
$
with $\delta(\mathbf{x})$ the Dirac-delta function. Such local two-body potentials 
are known to be well-suited for most of the cold-atom systems.

For instance, in the case of Hubbard-type tight-binding Hamiltonians with onsite 
interactions, Eq.~(\ref{eqn:Ugen}) can be written as
\begin{align}
U_{n'm'\mathbf{k'}}^{nm\mathbf{k}}(\mathbf{q}) = \sum_S U_S
&{n'}_{\mathbf{k'}+\frac{\mathbf{q}}{2},\uparrow S}^* 
{m'}_{-\mathbf{k'}+\frac{\mathbf{q}}{2},\downarrow S}^* \nonumber \\
&\times m_{-\mathbf{k}+\frac{\mathbf{q}}{2},\downarrow S}
n_{\mathbf{k}+\frac{\mathbf{q}}{2},\uparrow S},
\label{eqn:US}
\end{align}
where $S$ labels the basis sites in a unit cell, i.e., sublattices in the system, $U_S$ 
is the onsite interaction with the possibility of a sublattice dependence, and
$
n_{\mathbf{k} \sigma S} = \langle S | n \mathbf{k} \sigma \rangle
$
is the projection of the Bloch function onto the $S$th sublattice.
Thus Eq.~(\ref{eqn:alphagen}) reduces to
\begin{align}
\label{eqn:alphaS}
\alpha_{nm\mathbf{k}}^\mathbf{q} = &- \frac
{\sum_S U_S n_{\mathbf{k}+\frac{\mathbf{q}}{2},\uparrow S}^* 
m_{-\mathbf{k}+\frac{\mathbf{q}}{2},\downarrow S}^*}
{\varepsilon_{n, \mathbf{k}+\frac{\mathbf{q}}{2}, \uparrow} 
+ \varepsilon_{m, -\mathbf{k}+\frac{\mathbf{q}}{2}, \downarrow} - E_\mathbf{q}} \nonumber \\
& \times \frac{1}{N_c}\sum_{n'm'\mathbf{k'}} 
{m'}_{-\mathbf{k'}+\frac{\mathbf{q}}{2},\downarrow S}
{n'}_{\mathbf{k'}+\frac{\mathbf{q}}{2},\uparrow S} 
\alpha_{n'm'\mathbf{k'}}^\mathbf{q}.
\end{align}
This integral equation suggests that one can determine all possible $E_q$ solutions 
by representing Eq.~(\ref{eqn:alphaS}) as an eigenvalue problem in the $nm\mathbf{k}$ 
basis, i.e., the two-body problem reduces to finding the eigenvalues of an $N^2 \times N^2$ 
matrix for each $\mathbf{q}$. Alternatively, one can introduce a new parameter set
$
\beta_{S \mathbf{q}} = \sum_{nm\mathbf{k}} 
n_{\mathbf{k}+\frac{\mathbf{q}}{2},\uparrow S} 
m_{-\mathbf{k}+\frac{\mathbf{q}}{2},\downarrow S}
\alpha_{nm\mathbf{k}}^\mathbf{q},
$
and reduce the integral Eq.~(\ref{eqn:alphaS}) to a self-consistency relation
\begin{align}
\label{eqn:betaS}
\beta_{S \mathbf{q}} = - \frac{1}{N_c} \sum_{nm\mathbf{k} S'}
\frac{U_{S'}
n_{\mathbf{K} \uparrow S'}^* m_{-\mathbf{K'} \downarrow S'}^*
m_{-\mathbf{K'} \downarrow S} n_{\mathbf{K} \uparrow S}}
{\varepsilon_{n, \mathbf{k}+\frac{\mathbf{q}}{2}, \uparrow} 
+ \varepsilon_{m, -\mathbf{k}+\frac{\mathbf{q}}{2}, \downarrow} - E_\mathbf{q}} \beta_{S' \mathbf{q}},
\end{align}
where
$
\mathbf{K} = \mathbf{k}+\frac{\mathbf{q}}{2}
$
and 
$
\mathbf{K'} = \mathbf{k}-\frac{\mathbf{q}}{2}
$
are used as a shorthand notation. Thus, for a given $\mathbf{q}$, the two-body problem 
reduces to finding the roots of a nonlinear equation that is determined by setting the 
determinant of an $N_b \times N_b$ matrix to $0$. We illustrate these two approaches 
in the next section, where we focus on the experimentally more relevant case of a
sublattice-independent onsite interactions, and set $U_S = - U$ with $U \ge 0$ for the 
attractive case of interest in this paper.

\section{Bipartite Lattices}
\label{sec:tp}

For the sake of simplicity, below we consider a generic bipartite lattice with a two-point 
basis as a nontrivial illustration of our results, and denote its sublattices with 
$S = \{A, B\}$. In this case, the self-consistency equations can be combined to give
$
\begin{pmatrix}
M_{AA}^\mathbf{q} & M_{AB}^\mathbf{q} \\
M_{BA}^\mathbf{q} & M_{BB}^\mathbf{q}
\end{pmatrix}
\begin{pmatrix}
\beta_{A \mathbf{q}} \\ \beta_{B \mathbf{q}} 
\end{pmatrix}
= 0,
$
where the matrix elements are
\begin{align}
\label{eqn:MSS}
M_{SS}^\mathbf{q} &= 1 - \frac{U}{N_c} \sum_{nm\mathbf{k}} 
\frac{|n_{\mathbf{k}+\frac{\mathbf{q}}{2}, \uparrow S}|^2 |m_{-\mathbf{k}+\frac{\mathbf{q}}{2}, \downarrow S}|^2}
{\varepsilon_{n, \mathbf{k}+\frac{\mathbf{q}}{2}, \uparrow} 
+ \varepsilon_{m, -\mathbf{k}+\frac{\mathbf{q}}{2}, \downarrow} - E_\mathbf{q}}, \\
M_{AB}^\mathbf{q} &= - \frac{U}{N_c} \sum_{nm\mathbf{k}} 
\frac{n_{\mathbf{K} \uparrow B}^* m_{-\mathbf{K'} \downarrow B}^*
n_{\mathbf{K} \uparrow A} m_{-\mathbf{K'} \downarrow A}}
{\varepsilon_{n, \mathbf{k}+\frac{\mathbf{q}}{2}, \uparrow} 
+ \varepsilon_{m, -\mathbf{k}+\frac{\mathbf{q}}{2}, \downarrow} - E_\mathbf{q}},
\label{eqn:MAB}
\end{align}
with $M_{BA}^\mathbf{q} = M_{AB}^{\mathbf{q}*}$. Thus the nontrivial bound-state 
solutions require the condition 
$
\det \mathbf{M}_\mathbf{q} = M_{AA}^\mathbf{q} M_{BB}^\mathbf{q} - |M_{AB}^\mathbf{q}|^2 = 0
$ 
to be satisfied. In this paper we are interested in the time-reversal symmetric systems where 
$
n_{\mathbf{k} \uparrow S} = n_{-\mathbf{k} \downarrow S}^* 
\equiv n_{\mathbf{k} S} = \langle S | n \mathbf{k} \rangle.
$

In the presence of two sublattices, the one-body contributions to the Hamiltonian 
can be written as 
\begin{align}
\label{eqn:H0k}
H_0 = \sum_{\mathbf{k} \sigma} 
\begin{pmatrix}
c_{A \mathbf{k} \sigma}^\dagger & c_{B \mathbf{k} \sigma}^\dagger
\end{pmatrix}
( d_\mathbf{k}^0 \tau_0 + \mathbf{d_\mathbf{k}} \cdot \boldsymbol{\tau} )
\begin{pmatrix}
c_{A \mathbf{k} \sigma} \\ c_{B \mathbf{k} \sigma}
\end{pmatrix},
\end{align}
where $c_{S \mathbf{k} \sigma}$ annihilates a spin-$\sigma$ fermion in the 
$S$th sublattice with momentum $\mathbf{k}$, and $d_\mathbf{k}^0$ and 
$\mathbf{d_\mathbf{k}} = (d_\mathbf{k}^x, d_\mathbf{k}^y, d_\mathbf{k}^z)$
parametrize the most general Hamiltonian matrix in the sublattice basis.
Here $\tau_0$ is an identity matrix and $\boldsymbol{\tau} = (\tau_x, \tau_y, \tau_z)$ 
is a vector of Pauli spin matrices. The one-body dispersions 
$
\varepsilon_{s \mathbf{k} \uparrow} = \varepsilon_{s, -\mathbf{k}, \downarrow}
= \varepsilon_{s \mathbf{k}} 
$
are given by
\begin{align}
\label{eqn:esk}
\varepsilon_{s \mathbf{k}} = d_\mathbf{k}^0 + s d_\mathbf{k},
\end{align}
where $s = \pm$ denotes the upper and lower bands, and
$
d_\mathbf{k} = \sqrt{(d_\mathbf{k}^x)^2+(d_\mathbf{k}^y)^2+(d_\mathbf{k}^z)^2}
$
is the magnitude of $\mathbf{d_\mathbf{k}}$. The sublattice projections of the 
Bloch functions can be written as
\begin{align}
\label{eqn:Ask}
\langle A |s \mathbf{k} \rangle &= \frac{-d_\mathbf{k}^x + i d_\mathbf{k}^y}
{\sqrt{2d_\mathbf{k}(d_\mathbf{k} - s d_\mathbf{k}^z)}}, \\
\langle B |s \mathbf{k} \rangle &= \frac{d_\mathbf{k}^z - s d_\mathbf{k}}
{\sqrt{2d_\mathbf{k}(d_\mathbf{k} - s d_\mathbf{k}^z)}}.
\label{eqn:Bsk}
\end{align}
By plugging these expressions into Eqs.~(\ref{eqn:MSS}) and~(\ref{eqn:MAB}), we find
\begin{align}
\label{eqn:MAAq}
M_{AA}^\mathbf{q} &= 1 - \frac{U}{4N_c} \sum_{ss'\mathbf{k}} \frac{
\left(1 + s\frac{d_\mathbf{K}^z}{d_\mathbf{K}} \right)
\left(1 + s'\frac{d_\mathbf{K'}^z}{d_\mathbf{K'}} \right)}
{\varepsilon_{s, \mathbf{k}+\frac{\mathbf{q}}{2}} + \varepsilon_{s', \mathbf{k}-\frac{\mathbf{q}}{2}} - E_\mathbf{q}}, \\
\label{eqn:MBBq}
M_{BB}^\mathbf{q} &= 1 - \frac{U}{4N_c} \sum_{ss'\mathbf{k}} \frac{
\left(1 - s\frac{d_\mathbf{K}^z}{d_\mathbf{K}} \right)
\left(1 - s'\frac{d_\mathbf{K'}^z}{d_\mathbf{K'}} \right)}
{\varepsilon_{s, \mathbf{k}+\frac{\mathbf{q}}{2}} + \varepsilon_{s', \mathbf{k}-\frac{\mathbf{q}}{2}} - E_\mathbf{q}}, \\
\label{eqn:MABq}
M_{AB}^\mathbf{q} &= - \frac{U}{4N_c} \sum_{ss'\mathbf{k}} 
\frac{s\frac{d_\mathbf{K}^x - id_\mathbf{K}^y} {d_\mathbf{K}}
s'\frac{d_\mathbf{K'}^x + id_\mathbf{K'}^y} {d_\mathbf{K'}}}
{\varepsilon_{s, \mathbf{k}+\frac{\mathbf{q}}{2}} + \varepsilon_{s', \mathbf{k}-\frac{\mathbf{q}}{2}} - E_\mathbf{q}}.
\end{align}
Before proceeding with the numerical applications, next we show that these exact 
expressions are in perfect agreement with those of the Gaussian-fluctuation theory 
that is presented in Ref.~\cite{iskin20}.

To reveal a direct link between the variational approach to the two-body bound-state 
problem and the effective-action approach to the many-body pairing problem in 
the Gaussian approximation, first we consider the normal state with a vanishing 
saddle-point order parameters in the system, i.e., $\Delta_A = \Delta_B = 0$ for the 
sublattices. Then we substitute $\omega + 2\mu = E_\mathbf{q}$ after the analytical 
continuation of the Matsubara frequency $i\nu_\ell = \omega + i0^+$ of the pairs, and
take the zero-temperature limit. Within the Gaussian approximation, the fluctuation 
contribution to the thermodynamic potential can be written as
$
\Omega_G = \sum_\mathbf{q} 
\begin{pmatrix}
\Lambda_{T \mathbf{q}}^* & \Lambda_{R \mathbf{q}}^* 
\end{pmatrix}
\begin{pmatrix}
F_{TT}^\mathbf{q} & F_{TR}^\mathbf{q} \\
F_{RT}^\mathbf{q} & F_{RR}^\mathbf{q}
\end{pmatrix}
\begin{pmatrix}
\Lambda_{T \mathbf{q}} \\ \Lambda_{R \mathbf{q}}
\end{pmatrix},
$
where
$
\Lambda_{T \mathbf{q}} = (\Lambda_{A \mathbf{q}}+\Lambda_{B \mathbf{q}})/2
$
describes the total fluctuations and 
$
\Lambda_{R \mathbf{q}} = (\Lambda_{A \mathbf{q}}-\Lambda_{B \mathbf{q}})/2
$
describes the relative fluctuations. In Ref.~\cite{iskin20}, $\Lambda_{S q}$ is defined 
as the fluctuations of the complex Hubbard-Stratonovich field $\Delta_{S q}$ around 
the saddle-point order parameter $\Delta_S$ for the $S$th sublattice, 
i.e., $\Delta_{S q} = \Delta_S + \Lambda_{S q}$.
The matrix elements are reported as~\cite{iskin20}
\begin{align}
\label{eqn:FTT}
F_{TT}^\mathbf{q} &= \frac{1}{U} - \frac{1}{2N} \sum_{ss' \mathbf{k}} \frac{
1 + ss' \frac{d_\mathbf{K}^x d_\mathbf{K'}^x
+ d_\mathbf{K}^y d_\mathbf{K'}^y
+ d_\mathbf{K}^z d_\mathbf{K'}^z}
{d_{\mathbf{k}+\frac{\mathbf{q}}{2}}d_{\mathbf{k}-\frac{\mathbf{q}}{2}}}}
{\varepsilon_{s, \mathbf{k}+\frac{\mathbf{q}}{2}} + \varepsilon_{s', \mathbf{k}-\frac{\mathbf{q}}{2}} - E_\mathbf{q}}, \\
\label{eqn:FRR}
F_{RR}^\mathbf{q} &= \frac{1}{U} - \frac{1}{2N} \sum_{ss' \mathbf{k}} \frac{
1 - ss' \frac{d_\mathbf{K}^x d_\mathbf{K'}^x
+ d_\mathbf{K}^y d_\mathbf{K'}^y
- d_\mathbf{K}^z d_\mathbf{K'}^z}
{d_{\mathbf{k}+\frac{\mathbf{q}}{2}}d_{\mathbf{k}-\frac{\mathbf{q}}{2}}}}
{\varepsilon_{s, \mathbf{k}+\frac{\mathbf{q}}{2}} + \varepsilon_{s', \mathbf{k}-\frac{\mathbf{q}}{2}} - E_\mathbf{q}}, \\
\label{eqn:FTR}
F_{TR}^\mathbf{q} &= - \frac{1}{2N} \sum_{ss' \mathbf{k}} \frac{
s\frac{d_\mathbf{K}^z}{d_\mathbf{K}}
+ s'\frac{d_\mathbf{K'}^z}{d_\mathbf{K'}}
 - i ss' \frac{
d_\mathbf{K}^x d_\mathbf{K'}^y
-d_\mathbf{K}^y d_\mathbf{K'}^x}
{d_{\mathbf{k}+\frac{\mathbf{q}}{2}}d_{\mathbf{k}-\frac{\mathbf{q}}{2}}}}
{\varepsilon_{s, \mathbf{k}+\frac{\mathbf{q}}{2}} + \varepsilon_{s', \mathbf{k}-\frac{\mathbf{q}}{2}} - E_\mathbf{q}},
\end{align}
where $F_{RT}^\mathbf{q} = F_{TR}^{\mathbf{q}*}$. Here $N = 2N_c$ is the number of lattice
sites in the system, i.e., $N_b = 2$. We note that since the elements of $\mathbf{F}_\mathbf{q}$ 
and $\mathbf{M}_\mathbf{q}$ are related to each other through a unitary transformation, 
the condition 
$
\det \mathbf{F}_\mathbf{q} = F_{TT}^\mathbf{q} F_{RR}^\mathbf{q} - |F_{TR}^\mathbf{q}|^2 = 0
$
coincides precisely with $\det \mathbf{M}_\mathbf{q} = 0$.

\section{Numerical Application}
\label{sec:num}

\begin{figure} [htb]
\centerline{\scalebox{0.46}{\includegraphics{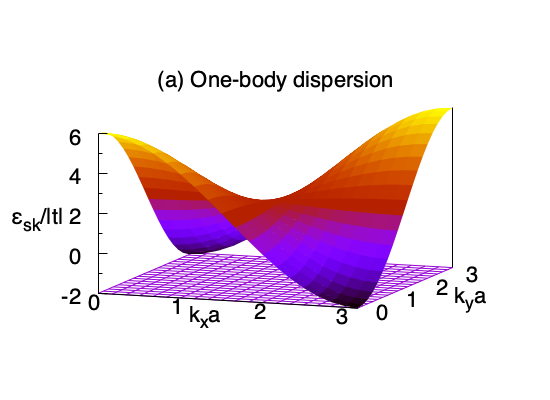}}}
\centerline{\scalebox{0.5}{\includegraphics{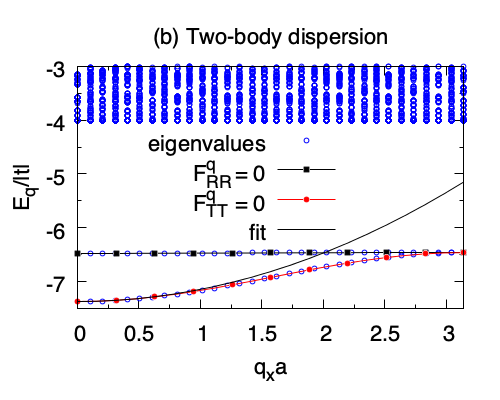}}}
\caption{\label{fig:U5} 
(a) One-body dispersion $\varepsilon_{s\mathbf{k}}$ is shown for the Mielke checkerboard 
lattice when the lower band is flat. The bands touch at the four corners of the 1st Brillouin zone.
(b) Two-body dispersion $E_\mathbf{q}$ is shown for $U = 5|t|$ as a function of $q_x a$
when $q_ya = 0$. The conditions $F_{RR}^\mathbf{q} = 0$ and $F_{TT}^\mathbf{q} = 0$ 
are in perfect agreement with the upper and lower branches, respectively. The quadratic 
expansion $E_\mathbf{q} = E_b + q^2/(2m_b)$ is an excellent fit for the lower branch in 
the small-$q$ limit.
}
\end{figure}

As a specific illustration of the theory, next we apply our generic results to study the two-body 
problem in a non-isolated flat band, i.e., a flat band that is in touch with others. In this context 
the Mielke checkerboard lattice in two dimensions is one of the simplest one to study since it 
exhibits a single flat band that is in touch with a single dispersive band at some $\mathbf{k}$ points. 
Such a lattice can be described by
$
d_\mathbf{k}^0 = -2t \cos(k_x a) \cos(k_y a),
$
$
d_\mathbf{k}^x = -2t \cos(k_x a) - 2t \cos(k_y a),
$
$
d_\mathbf{k}^y = 0,
$
and
$
d_\mathbf{k}^z = 2t \sin(k_x a) \sin(k_y a)
$
~\cite{iskin19b}.
Here $a$ is the lattice spacing between the nearest-neighbor sites of a square lattice, 
and the primitive vectors
$
\mathbf{b_1} = (\pi/a, -\pi/a) 
$
and
$
\mathbf{b_2} = (\pi/a, \pi/a)
$
determine the reciprocal lattice.
In this paper we let $t \to -|t|$ because it is advantageous to have the flat band as the 
lower one. This is because, no matter how weak $U$ is, the low-energy bound states 
that are most relevant to the presence of a flat band appear just below it,
i.e., they do not overlap with the one-body states. Thus the dispersive band
$
\varepsilon_{+,\mathbf{k}} = 2|t| + 4|t| \cos(k_x a) \cos(k_y a)
$
touches quadratically to the flat band
$
\varepsilon_{-,\mathbf{k}} = -2|t|
$
at the four corners of the 1st Brillouin zone
$
\mathbf{k} \equiv \{(\pm \pi/a, 0), (0, \pm \pi/a)\}.
$
A portion of the band structure is shown in Fig.~\ref{fig:U5}(a) for an extended zone.

For the two-body problem of interest in this paper, first we find all possible $E_\mathbf{q}$ 
values by solving the eigenvalue problem that is governed by Eq.~(\ref{eqn:alphaS}). 
The exact solutions are shown in Fig.~\ref{fig:U5}(b) for $U = 5|t|$ when $q_y a = 0$.
Note that all of the high-energy bound states have an instability towards a one-body 
decay in the $-4|t| \le E_\mathbf{q} \le 12 |t|$ region. For this reason we focus only on the 
low-energy states with $E_\mathbf{q} < -4|t|$. 
In Fig.~\ref{fig:U5}(b) there are two distinct bound-state branches appearing in the 
two-body problem. In contrast to the upper branch that appears nearly featureless in 
the shown scale, the lower one disperses quadratically with momentum in the small-$q$ 
limit. Given that our quadratic expansion $E_\mathbf{q} = E_b + q^2/(2m_b)$ 
is an excellent fit around $q = 0$, next we analyze both the offset $E_b < -4|t|$ of the 
lower branch and its effective mass $m_b > 0$ in greater detail.

\begin{figure} [htb]
\centerline{\scalebox{0.45}{\includegraphics{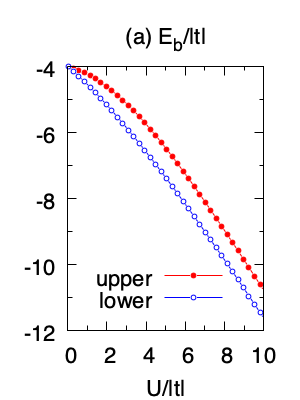} \includegraphics{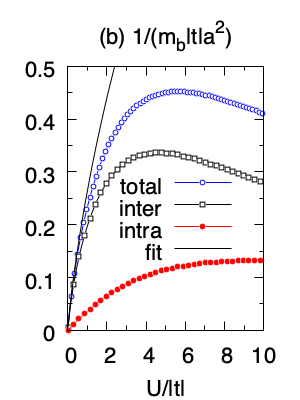}}}
\centerline{\scalebox{0.45}{\includegraphics{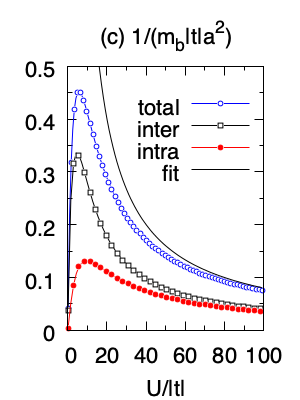}}}
\caption{\label{fig:Ebmb} 
(a) Lowest energy $E_b = E_{\mathbf{q} = \mathbf{0}}$ of the bound state 
is shown for the upper and lower branches as a function of $U$. 
(b) Inverse of the effective mass $m_b$ of the bound state is shown for the lower branch 
as a function of $U$ together with its intraband and interband contributions, where 
$1/m_b = 1/m_b^\mathrm{intra} + 1/m_b^\mathrm{inter}$. 
(c) Same as in (b) but with a larger region.
Here $m_b = 5\pi/[U a^2\ln(64|t|/U)]$ and $m_b = U/(8 a^2 t^2)$ fits very well in 
the small- and large-$U$ limits, respectively.
}
\end{figure}

For this purpose, first we note in Fig.~\ref{fig:U5}(b) that the conditions 
$F_{RR}^\mathbf{q} = 0$ and $F_{TT}^\mathbf{q} = 0$ are in perfect agreement with 
the upper and lower branches, respectively. This is because the coupling term 
$F_{TR}^\mathbf{q}$ integrates to $0$ when $q_x = 0$ and/or $q_y = 0$.
Then, in contrast to Eq.~(\ref{eqn:alphaS}), we note that Eqs.~(\ref{eqn:FTT}) 
and~(\ref{eqn:FRR}) offer an analytically tractable approach. For instance one can 
determine both $E_b$ and $m_b$ of the lower branch by substituting
$
E_\mathbf{q} = E_b + \sum_{ij} q_i (\mathbf{m_b^{-1}})_{ij} q_j / 2
$
in Eq.~(\ref{eqn:FTT}), and expanding the condition $F_{TT}^\mathbf{q} = 0$ up to 
second order in $\mathbf{q}$. Here $(\mathbf{m_b^{-1}})_{ij}$ corresponds to the $ij$th 
element of the inverse of the effective-mass tensor $\mathbf{m_b}$ of the lower branch.
Thus the condition $F_{TT}^\mathbf{0} = 0$ for the zeroth-order term leads to a 
closed-form expression
\begin{align}
\label{eqn:Eb}
1 = \frac{U}{N} \sum_{s \mathbf{k}} \frac{1}{2 \varepsilon_{s \mathbf{k}} - E_b}
\end{align}
for the $E_b$ of the lower branch. Note that the familiar one-band result is recovered 
by Eq.~(\ref{eqn:Eb}), after setting $d_\mathbf{k} = 0$ in the one-body dispersion 
shown in Eq.~(\ref{eqn:esk}). Similarly the condition $F_{RR}^\mathbf{0} = 0$ gives 
an expression for the $E_b$ of the upper branch. 
In Fig.~\ref{fig:Ebmb}(a) we show $E_b$ for both the upper and lower branches as a 
function of $U$. For the lower branch of main interest here, we find that 
$E_b = -4|t| - U/2$ is an excellent fit in the small-$U$ limit but it approaches to 
$E_b = -4|t| - U$ in the large-$U$ limit.

While the condition 
$
\partial F_{TT}^\mathbf{q}/\partial q_i |_{\mathbf{q} = \mathbf{0}} = 0
$ 
for the first-order term is always satisfied, the condition 
$
\partial^2 F_{TT}^\mathbf{q}/(\partial q_i\partial q_j) |_{\mathbf{q} = \mathbf{0}} = 0
$ 
for the second-order term leads to a closed-form expression 
$
(\mathbf{m_b^{-1}})_{ij} = (\mathbf{m_b^{-1}})_{ij}^\mathrm{intra} 
+ (\mathbf{m_b^{-1}})_{ij}^\mathrm{inter}
$
for the effective-mass tensor, where
\begin{align}
\label{eqn:intra}
(\mathbf{m_b^{-1}})_{ij}^\mathrm{intra} &= \frac{1}{2} \frac
{
\sum_{s \mathbf{k}} \frac{
\partial^2 \varepsilon_{s \mathbf{k}} / (\partial k_i \partial k_j)
}{(2\varepsilon_{s \mathbf{k}} - E_b)^2}}
{\sum_{s \mathbf{k}} \frac{1}{(2\varepsilon_{s \mathbf{k}} - E_b)^2}}, \\
(\mathbf{m_b^{-1}})_{ij}^\mathrm{inter} &= - 2 \frac
{
\sum_{s \mathbf{k}} \frac{
s d_\mathbf{k} g_\mathbf{k}^{ij}}
{(2d_\mathbf{k}^0 - E_b) (2\varepsilon_{s \mathbf{k}} - E_b)}
}
{\sum_{s \mathbf{k}} \frac{1}{(2\varepsilon_{s \mathbf{k}} - E_b)^2}},
\label{eqn:inter}
\end{align}
are the so-called intraband and interband contributions, respectively. Here
$
2 g_\mathbf{k}^{ij} = \partial (\mathbf{d}_\mathbf{k}/d_\mathbf{k}) \partial k_i
\cdot 
\partial (\mathbf{d}_\mathbf{k}/d_\mathbf{k}) \partial k_j
$
is precisely the quantum-metric tensor of the Bloch states~\cite{torma17a, iskin19b, iskin20}. 
It is truly delightful to note that the expressions Eqs.~(\ref{eqn:intra}) and~(\ref{eqn:inter}) 
are formally equivalent to the ones reported in the recent literature in an entirely 
different but a related context, i.e., the effective-mass tensor of the Cooper pairs 
in the presence of helicity bands that is induced by spin-orbit coupling~\cite{iskin18a}. 
In particular they suggest that while the intraband processes depend only on 
the one-body band structure, the interband ones are controlled by the 
quantum geometry of the Bloch states. In addition the familiar one-band result is 
recovered merely by Eq.~(\ref{eqn:intra}), after setting $d_\mathbf{k} = 0$ in the 
one-body dispersion shown in Eq.~(\ref{eqn:esk}). This leads not only to 
$
(\mathbf{m_b^{-1}})_{ij}^\mathrm{inter} = 0
$
but also to
$
(\mathbf{m_b^{-1}})_{ij}^\mathrm{intra} = \delta_{ij}/(2m)
$
for the one-body dispersion that is quadratic in $k$, e.g., 
$d_\mathbf{k}^0 = \varepsilon_0 + k^2/(2m)$, where $\delta_{ij}$ is the Kronecker-delta. 

For the specific case of a Mielke checkerboard lattice, $\mathbf{m_b}$ turns out to be 
a diagonal matrix with isotropic elements, leading to
$
1/m_b = 1/m_b^\mathrm{intra} + 1/m_b^\mathrm{inter},
$ 
and they are shown in Fig.~\ref{fig:Ebmb}(b) as a function of $U$. By the trial and error 
approach, we find that
$
m_b = 5\pi/[U a^2\ln(64|t|/U)]
$ 
fits very well in the small-$U$ limit. Since the effective intraband mass of the one-body 
dispersion diverges for the flat band to begin with, we note that $U \ne 0$ is responsible 
for $m_b \ne \infty$ through the interband processes with the dispersive band, e.g.,
it can be shown that
$
(\mathbf{m_b^{-1}})_{ij}^\mathrm{inter} \approx \frac{U}{N} \sum_\mathbf{k} g_\mathbf{k}^{ij}
[1 - U/(4\varepsilon_{+, \mathbf{k}} - 2E_b)]
$
in the $U \to 0^+$ limit. Here $\frac{1}{N_c} \sum_\mathbf{k} g_\mathbf{k}^{ij}$ diverges 
by itself due to the touching points, and the second term is crucial for producing 
a finite effective mass in the Mielke flat band, i.e., it cancels precisely those diverging points.
Thus our calculation reveals the quantum-geometric mechanism that gives rise to a 
finite $m_b$ in the $U \to 0^+$ limit as long as $U$ is nonzero. 
However, away from the small-$U$ limit, Fig.~\ref{fig:Ebmb}(b) shows that the intraband 
processes within the dispersive band also give a similar contribution. The physical 
mechanism is known to be very different in the large-$U$ 
limit~\cite{ohashi08, wouters06, valiente08}, where the tunneling of the bound state 
is possible only through virtual dissociation of the pair, and this leads to 
$m_b \sim U/(8 a^2 t^2)$ as shown in Fig.~\ref{fig:Ebmb}(c).

In particular to the small-$U$ limit, we would like to emphasize that our generic result 
$
m_b \propto \mathcal{A}/[U \ln (\mathcal{B}/U)]
$ 
for the non-isolated flat bands is in distinct contrast with that 
$
m_b \propto \mathcal{A}/U
$ 
of the isolated ones~\cite{torma18}, where $\mathcal{A}$ and $\mathcal{B}$ are real 
constants depending on the lattice structure. To be more precise, it was found that the 
quadratic expansion of $E_\mathbf{q}$ works very well for some isolated flat bands 
with an offset
$
E_b = -U/N_b
$
defined from the flat band and an effective-mass tensor
$
(\mathbf{m_b^{-1}})_{ij} = \frac{U}{N} \sum_\mathbf{k} g_\mathbf{k}^{ij}
$
in the small-$U$ limit~\cite{torma18}. Here $g_\mathbf{k}^{ij}$ is the corresponding 
quantum-metric tensor of the Bloch states in the flat band in the presence of other flat and/or 
dispersive bands. In comparison to the intraband contribution of Eq.~(\ref{eqn:intra}) 
for a non-isolated flat band, there is no such contribution for an isolated flat band 
in the small-$U$ limit due to the presence of a band gap between the flat band 
and others. However, we again note that $U \ne 0$ is fully responsible for 
$m_b \ne \infty$ through merely the interband processes with the rest of the Bloch 
states in the system.

\section{Conclusion}
\label{sec:conc}

In summary, above we constructed a variational approach to study the two-body bound-state 
problem in a generic multi-band lattice, and gave a detailed account of bipartite lattices 
with an onsite interaction that manifest time-reversal symmetry. For this case we 
showed that the lowest-energy bound states disperse quadratically with momentum, 
whose effective-mass tensor has two physically distinct contributions coming from (i) the 
intraband processes that depend only on the one-body dispersion and (ii) the interband 
processes that also depend on the quantum-metric tensor of the underlying Bloch states. 
In particular we applied our theory to the Mielke checkerboard lattice for its simplicity, 
and revealed how the interband processes help produce a finite effective mass for the 
bound states in a non-isolated flat band. As an outlook, our theory can be extended to 
the non-isolated flat bands of Kagome and Lieb lattices that have recently been realized 
in a number of physical systems~\cite{jo12, nakata12, li18, diebel16, kajiwara16, ozawa17}.

\begin{acknowledgments}
The author acknowledges funding from T{\"U}B{\.I}TAK Grant No. 1001-118F359.
\end{acknowledgments}

\end{document}